\documentclass[a4paper]{article}

\usepackage[utf8]{inputenc}
\usepackage[T1]{fontenc}

\usepackage{amsmath}
\usepackage{amssymb}
\usepackage{hyperref}

\DeclareMathOperator{\diver}{div}

\def\cV{{\mathcal V}} 
\def\cG{{\mathcal G}}
\def\cE{{\mathcal E}} 
\def\cP{{\mathcal P}}
\def\R{\mathbb{R}}
\def\C{\mathbb{C}}
\def\d{\textup{d}}
\DeclareMathOperator{\e}{{\bold e}}

{\left(\begin{smallmatrix}}%            begin code
{\end{smallmatrix}\right)}%             end code

\renewcommand{\theta}{\vartheta}
\renewcommand{\phi}{\varphi}

\usepackage{pgf,tikz,pgfplots}
\pgfplotsset{compat=1.14}
\usepackage{mathrsfs}
\usetikzlibrary{arrows}

\usepackage[amsmath,amsthm,thmmarks]{ntheorem}

\theoremstyle{plain}
\newtheorem{definition}{Definition}[section]
\newtheorem{theorem}[definition]{Theorem}
\newtheorem{proposition}[definition]{Proposition}
\newtheorem{lemma}[definition]{Lemma}

\newtheorem{example}[definition]{Example}
\newtheorem{remark}[definition]{Remark}

\title{A Calder\'on type inverse problem for tree graphs
 \thanks{The second author gratefully acknowledges financial support by 
the grant no.\ 2018-04560 of the Swedish Research Council (VR).}
        }

\author{Hannes Gernandt\thanks{Institut für Mechanik und Meerestechnik, TU Hamburg, Ei\ss endorfer Stra\ss e 42, 21073 Hamburg, Germany ({\tt hannes.gernandt@tuhh.de}).}
        \and Jonathan Rohleder \thanks{ Matematiska institutionen, Stockholms universitet, 106 91 Stockholm, Sweden ({\tt jonathan.rohleder@math.su.se}).}}

\begin{document}

\maketitle

\begin{abstract}
We study the inverse problem of recovering a tree graph together with the 
weights on its edges (equivalently a metric tree) from the knowledge of the Dirichlet-to-Neumann matrix associated with the Laplacian. We prove an explicit formula which relates this matrix to the pairwise weighted distances of the leaves of the tree 
and, thus, allows to recover the weighted tree. This result can be viewed 
as a counterpart of the Calder\'on problem in the analysis of PDEs. In contrast to earlier results on inverse problems for metric graphs, we only assume knowledge of the Dirichlet-to-Neumann matrix for a fixed energy, not of a whole matrix-valued function. 
\end{abstract}
%\textit{Keywords:}  quantum graphs,\ direct sum operators,\  dirac operators, boundary triplets\vspace{0.5cm}\\

%\textit{MSC 2010:} 15A22, 15A18, 47A55

%\begin{keywords}

%\end{keywords}

%\begin{AMS}

%\end{AMS}

%\pagestyle{myheadings}
%\thispagestyle{plain}
%\markboth{\textsc{H. Gernandt and C. Trunk}}{\textsc{Eigenvalue placement for regular matrix pencils}}

\section{Introduction}

Calder\'on's classical problem from Electrical Impedance Tomography consists in recovering, if possible, the isotropic conductivity of an inhomogeneous body uniquely from applying voltages to the surface of the body and 
measuring the corresponding current flux through the surface, see~\cite{C80}. In mathematical terms the relation between voltage and current is given by the so-called Dirichlet-to-Neumann map $M_{\Omega, \gamma}$ on the 
boundary of a Euclidean domain $\Omega$ with a positive conductivity $\gamma : \Omega \to \R$. It is defined by the relation
\begin{align*}
 M_{\Omega, \gamma} u |_{\partial \Omega} = \partial_\nu u |_{\partial \Omega},
\end{align*}
where $u$ satisfies $\diver \gamma \nabla u = 0$ in $\Omega$ and $u |_{\partial \Omega}$ and $\partial_\nu u |_{\partial \Omega}$ denote the trace and the derivative with respect to the unit normal, respectively, of 
$u$ on the boundary. This problem was proven to be uniquely solvable under reasonable regularity assumptions; see, e.g.,~\cite{AP06,N88,N96,SU87}.

In the context of discrete graphs, the inverse conductivity problem may be read in the following way: given a discrete graph $\cG$, consisting of 
a finite vertex set $\cV$ and a finite edge set $\cE$, conductivities on the edges correspond to positive edge weights $w_e$, $e \in \cE$. Moreover, the Dirichlet-to-Neumann matrix $M_{\cG}$ for the weighted graph $\cG$ 
and a selected set $\partial \cG = \{v_1, \dots, v_k\}$ of vertices (considered as the boundary of the graph) may be defined by the relation
\begin{align*}
 \left( M_{\cG} \phi \right) (v_i) = \frac{\partial u_\phi}{\partial \rm n} (v_i), \quad i = 1, \dots, k,
\end{align*}
for any function $\phi : \partial \cG \to \C$; here $u_\phi: \cV\to \C$ is the harmonic extension of $\phi$, i.e.\ it satisfies $L (\cG) u_\phi (v) = 0$ at each vertex $v \notin \partial \cG$ for the discrete weighted 
Laplacian $L (\cG)$, and $\frac{\partial u_\phi}{\partial \rm n} (v_i)$ is the normal derivative of $u_\phi$ at $v_i$; see Section \ref{sec:prelim} for details. Hence the exact analogue of Calder\'on's problem consists in recovering the weights $w_e$ from the knowledge of $M_\cG$ for a properly chosen boundary $\partial \cG$. This problem is not new and has been solved 
for certain prescribed graph topologies such as rectangular grids \cite{CurtMorr90,CurtMorr91}, certain planar graphs \cite{CurtInge98}, or certain circular ``spiderweb'' networks \cite{CurtMooe94}, or under additional assumptions and in related settings \cite{AIM18,BGG16,BereChun05,CGHS17}. 
Especially, it has been solved for tree graphs if $\partial \cG$ consists 
of all leaves, i.e.\ all vertices of degree one, see \cite[Theorem~2]{CurtInge98}.

The aim of this note is, however, to go one step further and to recover simultaneously the tree graph itself and the weights of all edges from the Dirichlet-to-Neumann matrix on the leaves of the tree. The main idea is to compute the weighted distance (also called \emph{resistance distance}, see, e.g., \cite{KleiRand93}) of each pair of leaves from the Dirichlet-to-Neumann map $M_\cG$. In more detail, let $v_1, \dots, v_k$ be the leaves of $\cG$ and denote by $\d_\cG (\cdot, \cdot)$ the weighted distance on $\cG$; see Section \ref{sec:discrete} below. For any $i_0 \in 
\{1, \dots, k\}$ we denote by $S_{i_0} \in \R^{k \times k}$ the matrix obtained from $M_\cG$ by replacing the $i_0$-th diagonal entry by one and all other entries in the $i_0$-th row and column by zero. In our main result, Theorem \ref{thm:main}, we prove that $S_{i_0}$ is invertible and that the identity
\begin{align}\label{eq:distances}
 \left\langle S_{i_0}^{-1} \e_j, \e_j \right\rangle = \d_\cG (v_j, v_{i_0})
\end{align}
holds for each $j \in \{1, \dots, k\}$ with $j \neq i_0$. Thus the distance of each pair of leaves can be read off explicitly from the Dirichlet-to-Neumann matrix. It is well-known that a weighted tree can be recovered completely (up to vertices of degree two) from this information, see, e.g., \cite{HakiYau65}. Therefore the formula \eqref{eq:distances} allows to recover the connectivity of the tree as well as its edge weights from the Dirichlet-to-Neumann matrix $M_\cG$ at the same time. In particular, if $\cG_1$ and $\cG_2$ are two weighted trees which have the same number of leaves and satisfy $M_{\cG_1} = M_{\cG_2}$ then $\cG_1 = \cG_2$ up 
to vertices of degree two, and the two trees have the same edge weights. We complement this result in Section \ref{sec:examples} by a few examples 
which rule out several naive generalizations, e.g.\ to graphs with cycles 
or to partial data inverse problems. 

The main result of this note has also a natural interpretation in the language of differential operators on metric graphs, so-called quantum graphs (see, e.g., \cite{BK13}), and related inverse problems were considered in recent years. Given a finite metric graph~$\Gamma$ and an electric potential $q$ on $\Gamma$, one defines the Dirichlet-to-Neumann matrix $M_{\Gamma, q} (\lambda)$ for the differential equation $- f'' + q f = \lambda f$ on $\Gamma$ for suitable $\lambda \in \C$ by the equation
\begin{align*}
 M_{\Gamma, q} (\lambda) f_\lambda |_{\partial \Gamma} = \partial f_\lambda |_{\partial \Gamma},
\end{align*}
where $f_\lambda$ satisfies $- f_\lambda'' + q f_\lambda = \lambda f_\lambda$ inside the edges of $\Gamma$ and standard (continuity--Kirchhoff) matching conditions on all vertices that do not belong to the set of leaves $\partial \Gamma$; here $f_\lambda |_{\partial \Gamma}$ and $\partial f_\lambda |_{\partial \Gamma}$ are the vectors of boundary evaluations of 
$f_\lambda$ and its derivative, respectively. The function $\lambda \mapsto M_{\Gamma, q} (\lambda) \in \C^{k \times k}$ is a matrix-valued Herglotz--Nevanlinna function with a discrete set of poles on the real axis. In 
recent years the inverse problem of recovering the metric graph $\Gamma$ and the potential $q$ on it from the knowledge of the matrix function $\lambda \mapsto M_{\Gamma, q} (\lambda)$ has received a lot of attention. It was solved for the case that $\Gamma$ is a tree first in the equilateral case in~\cite{BW05,Y05} and later for arbitrary edge lengths in~\cite{AK08}; see also~\cite{CW09,FY07,R15,Y12} for related results. On the other 
hand, it is not uniquely solvable for more general graphs if not further additional data is provided, see~\cite{EK11,K13}. It can be seen easily (see Section \ref{sec:QG} below) that in the potential-free case, $q = 0$ identically, one has 
\begin{align*}
 M_{\Gamma, 0} (0) = M_\cG
\end{align*}
if $\cG$ is the underlying discrete graph of $\Gamma$, equipped with the edge weights $w_e = 1 / L (e)$, where $L (e)$ are the edge lengths in the metric graph. We point out that in the quantum graphs literature so far only inverse problems based on the knowledge of the matrix function $\lambda \mapsto M_{\Gamma, q} (\lambda)$ were considered, while we require only knowledge of its value at $\lambda = 0$ in the present note. Thus the present result also complements the literature on inverse problems for quantum graphs.

\section{Preliminaries}\label{sec:prelim}

Throughout this note, $\cG$ denotes a finite graph consisting of a finite 
set $\cV = \cV (\cG)$ of vertices and a finite set $\cE = \cE (\cG)$ of edges, together with edge weights $w_e > 0$, $e \in \cE$. For each $v \in \cV$ we denote by $\deg v$ its degree and by $\cE (v)$ the set of all 
edges incident to $v$. Moreover, we denote by $\partial \cG$ a selected set of vertices, called {\em boundary}, and call the corresponding vertices {\em boundary vertices}; later $\cG$ will be a tree, i.e.\ a graph without cycles, and $\partial \cG$ will consist of the vertices of degree one, but for now we allow a more general setting. Each vertex which is not a 
boundary vertex is called {\em interior vertex}. In the following we assume for simplicity that $\cG$ has no multiple edges and no loops. We assume also that $\cG$ is \textit{connected}, i.e.\ for any two vertices there 
exists a path connecting them. We denote the vertices of $\cG$ by $v_1, \dots, v_n$ and assume that $v_1, \dots, v_k$ are those vertices which belong to $\partial \cG$, where $1 \leq k = |\partial \cG| \leq n$. 

Consider the discrete Laplacian $L (\cG) \in \R^{n\times n}$ on $\cG$ given by
\begin{align*}
 \big( L (\cG) \big)_{i, j} = \begin{cases} - w_e & \text{if}~e~\text{connects}~v_i~\text{and}~v_j, i \neq j,\\ 0 & \text{if}~v_i, v_j~\text{are 
not adjacent}, \\ \sum_{e \in \cE (v_i)} w_e & \text{if}~i = j.  \end{cases}
\end{align*}
Equivalently, $L (\cG)$ acts on $u : \cV \to \C$ via
\begin{align*}
 \big( L (\cG) u \big) (v_i) = \sum_{e \in \cE (v_i) \cap \cE (v_j)} w_e \big( u (v_i) - u (v_j) \big), \quad i = 1, \dots, n.
\end{align*}
The operator $L (\cG)$ is symmetric and nonnegative, and its kernel consists of the constant vectors; for more details we refer the reader to the classical monograph \cite{C97}. 

Furthermore, for $u : \cV \to \C$ we define a {\em normal derivative} at any boundary vertex $v_i \in \partial \cG$ by
\begin{align*}
 \frac{\partial u}{\partial \rm n} (v_i) = \sum_{e \in \cE (v_i) \cap \cE (v_j)} w_e \big( u (v_i) - u (v_j) \big).
\end{align*}
Note that if $\deg v_i = 1$ then the sum in the definition of $\frac{\partial u}{\partial \rm n} (v_i)$ consists of only one summand. 

Next, observe that for each $\phi : \partial \cG \to \C$ there exists a unique function $u_\phi : \cV \to \C$ such that $(L (\cG) u_\phi) (v) = 0$ for all $v \in \cV \setminus \partial \cG$ and $u |_{\partial \cG} = 
\phi$; $u_\phi$ is called {\em harmonic extension} of $\phi$. We can therefore define the Dirichlet-to-Neumann matrix as follows.

\begin{definition}\label{def:DN}
Let $\cG$ be a weighted graph and let $v_1, \dots, v_k$ be the vertices belonging to its boundary $\partial \cG$. The {\em Dirichlet-to-Neumann matrix} corresponding to the Laplacian on $\cG$ is the $k \times k$-matrix $M_\cG$ that satisfies %\textbf{man muss schon eine ordnung der Randecken 
%festlegen...}
\begin{align*}
 M_\cG \begin{pmatrix}\phi(v_1)\\ \vdots \\ \phi(v_k)\end{pmatrix} := \begin{pmatrix} \frac{\partial u_\phi}{\partial \rm n}(v_1)\\ \vdots \\ \frac{\partial u_\phi}{\partial \rm n}(v_k)\end{pmatrix} ,
\end{align*}
where for each $\phi : \partial \cG \to \C$, $u_\phi$ is the corresponding harmonic extension.
\end{definition} 

It can be checked easily that the matrix $M_\cG$ is symmetric and nonnegative. In the special case that $\cG$ consists of only two vertices and one edge connecting them, one has $M_{\cG}=L(\cG)$. According to the partition of the vertices into the boundary vertices $v_1, \dots, v_k\in\partial \cG$ and the interior vertices $v_{k + 1}, \dots, v_n\in\cV\setminus\partial\cG$, we can write the discrete Laplacian as a block matrix 
\begin{align}\label{eq:decomp}
 L (\cG) = \begin{pmatrix} \widehat D & - B^\top \\ - B & \widehat L \end{pmatrix}.
\end{align}
It is well known, see, e.g., \cite[Theorem 3.2]{CurtInge98}, that $\widehat L$ is invertible and that the Dirichlet-to-Neumann matrix $M_{\cG}$ coincides with the Schur complement of $\widehat L$ in $L (\cG)$, i.e.\ 
\begin{align}
\label{DtoN_schur}
 M_\cG = \widehat D - B^\top \widehat L^{- 1} B.
\end{align}

\section{Reconstruction of a weighted tree from the Dirichlet-to-Neumann matrix}\label{sec:discrete}

In this section we prove the main result of this note. Throughout this section all graphs are trees, and we choose
\begin{align*}
 \partial \cG = \left\{ v \in \cV : \deg v = 1 \right\}.
\end{align*}

The proof of the main result requires some preparation. One of its main ingredients will be the following lemma, which can be found in~\cite[Proposition~1]{Kirkland1996}. Here the \textit{reduced Laplacian} $L_v (\cG)$ appears, which by definition is the matrix obtained from $L (\cG)$ by removing the line and column that correspond to the vertex $v$. 

\begin{lemma}
\label{lem:Kirkland}
Let $\cG$ be a finite weighted tree. Then for any vertex $v \in \cV$ the entry $(i,j)$ of the inverse $L_v (\cG)^{-1}$ of the reduced Laplacian equals $\sum_{e \in P_{i, j}} \frac{1}{w_e}$, where $P_{i, j}$ is the set of edges that are on both the path from $v_i$ to $v$ and the path from $v_j$ to $v$.
\end{lemma}

For the formulation of our main result we recall the following definition.
\begin{definition}
Two weighted graphs $\cG_1, \cG_2$ are called {\em equal}, $\cG_1 = \cG_2$, if $|\cV (\cG_1)| = |\cV (\cG_2)| =: n$ and there exists a permutation matrix $P \in \R^{n\times n}$ with $L (\cG_2) = P^\top L (\cG_1) 
P$. Moreover, $\cG_1$ and $\cG_2$ are called {\em equal up to vertices of 
degree two} if they are equal after removing each vertex $v$ of degree two and replacing the two edges $e_1, e_2$ incident to $v$ by one edge $e$ whose weight $w_e$ satisfies
%\begin{align*}
$w_e^{-1} = w_{e_1}^{-1} + w_{e_2}^{-1}$.
%\end{align*}
\end{definition}

In the following we consider distances of vertices in a weighted tree~$\cG$ with edge weights $w_e$, $e \in \cE$. If $\cP$ is the unique path in $\cG$  connecting two vertices $v_i$ and $v_j$ consisting of the edges $e_1, \dots, e_m$, then we define
\begin{align*}
 \d_{\cG} (v_i, v_j) = \sum_{k = 1}^{m} \frac{1}{w_{e_k}}.
\end{align*}
In the context of inverse conductivity problems oftentimes $w_{e_i}$ is the conductivity and therefore $w_{e_i}^{-1}$ can be interpreted as a resistance. In this setting the distance  $\d_{\cG} (v_i, v_j)$ coincides with the \emph{resistance distance} introduced in \cite{KleiRand93}.

Below we recall that trees are uniquely determined up to vertices of degree two from the pairwise distances between the boundary vertices; for a proof see, e.g.,~\cite[Theorem 3]{HakiYau65}. 

\begin{proposition}\label{prop:lengths}
Let $\cG_1, \cG_2$ be two weighted trees with boundaries
\begin{align*}
 \partial \cG_j = \big\{v_1^j, \dots, v_k^j \big\}, \quad j = 1, 2,
\end{align*}
where $k = |\partial \cG_1| = |\partial \cG_2|$. Assume that the pairwise distances of boundary vertices in $\cG_1$ and $\cG_2$ coincide, i.e.
\begin{align*}
 \d_{\cG_1} (v_i^1, v_m^1) = \d_{\cG_2} (v_i^2, v_m^2)
\end{align*}
holds for $i, m = 1, \dots, k$. Then $\cG_1 = \cG_2$ up to vertices of degree two.
\end{proposition}
The construction of the tree from the pairwise distances between boundary 
vertices is straightforward and e.g.\ described in \cite[Lemma 2.16]{Ruec11}. The key observation is that the pairwise distances between a triplet 
of boundary vertices can be used to determine the distance to a branching 
point, where the paths from one boundary vertex to the other two vertices 
split. 

We are now in the position to prove the main result of this note. 

\begin{theorem}\label{thm:main}
Let $\cG$ be a weighted tree with boundary $\partial \cG = \{v_1, \dots, v_k\}$ and let $M_\cG$ be the corresponding Dirichlet-to-Neumann matrix 
in Definition \ref{def:DN}. Denote by $S_{i_0} \in \R^{k \times k}$ the matrix obtained from $M_\cG$ by replacing the $i_0$-th diagonal entry by one and all other entries in the $i_0$-th row and column by zero. Then $S_{i_0}$ is invertible and for each $j \neq i_0$
\begin{align}\label{eq:Hauptgewinn}
 \big\langle S_{i_0}^{-1} \e_j, \e_j \big\rangle = \d_\cG (v_j, v_{i_0}),
\end{align}
i.e.\ the $j$-th diagonal entry of $S_{i_0}^{-1}$ equals the distance between the boundary vertices $v_{i_0}$ and $v_j$. In particular, the tree $\cG$ together with its edge weights can be recovered uniquely, up to vertices of degree two, from the Dirichlet-to-Neumann matrix.
\end{theorem}

\begin{proof}
{\em \underline{Step~1:}} In this step we calculate an explicit expression for $S_{i_0}^{-1}$. We make use of the block matrix decomposition \eqref{eq:decomp} of $M_\cG$. Denote by~$U$ the matrix obtained from $B$ by replacing its $i_0$-th column by zero. Moreover, let $\dot \cG$ denote the weighted tree obtained from $\cG$ by removing all boundary vertices (and their incident edges) and by $L (\dot \cG)$ the corresponding discrete Laplacian on $\dot \cG$. Note that for $l = 1, \dots, n - k$
\begin{align}\label{eq:Btop}
 B^\top \e_l = \sum w_{e_j} \e_j,
\end{align}
where the sum is taken over all $j$ such that the boundary vertex $v_j$ is adjacent to the $l$-th interior vertex $v_{k + l}$, and $e_j$ is the edge incident to $v_j$. As a consequence, the matrix $B \widehat D^{-1} B^\top \in \R^{(n - k) \times (n - k)}$ is diagonal and its $l$-th diagonal entry equals
\begin{align*}
 \big\langle \widehat D^{-1} B^\top \e_l, B^\top \e_l \rangle = \sum w_{e_j},
\end{align*}
with the sum being taken over the same $j$ as for~\eqref{eq:Btop}.
 Therefore the matrix $\widehat L$ in the decomposition~\eqref{eq:decomp} 
can be written
\begin{align}\label{eq:zumEinen}
 \widehat L = L (\dot \cG) + B \widehat D^{-1} B^\top = L (\dot \cG) + w_{e_{i_0}} \e_{l_0} \e_{l_0}^\top + U \widehat D^{-1} U^\top,
\end{align}
where $e_{i_0}$ is the edge incident to the boundary vertex $v_{i_0}$ and 
$l_0$ is such that $v_{k + l_0}$ is the vertex at the other end of $e_{i_0}$. 

Let us write $L := L (\dot \cG) + w_{e_{i_0}} \e_{l_0} \e_{l_0}^\top$, so that~\eqref{eq:zumEinen} can be rewritten 
\begin{align}\label{eq:zumAnderen}
 \widehat L = L + U \widehat D^{-1} U^\top.
\end{align}
The matrix $L$ is invertible since $L(\dot \cG)$ is positive semi-definite and its kernel is spanned by the all-ones vector which is linearly independent of $\e_{l_0}$. It follows that in fact $L$ is positive definite, and since $U \widehat D^{-1} U^\top$ is nonnegative, also $L + U \widehat D^{-1} U^\top$ is positive definite and, in particular, invertible. 
%%%%%%%%%%%%%%%% NEW PART. %%%%%%%%%%%%%%%%%
Therefore~\eqref{eq:zumAnderen} together with the Sherman-Morrison-Woodbury formula implies 
\begin{align*}
 \big( \widehat D - U^\top \widehat L^{-1} U \big)^{-1} 
%  &= \widehat D^{-1} - \widehat D^{-1} U^\top \widehat L^{-1/2} (I-\widehat L^{-1/2} U\widehat D^{-1} U^\top \widehat L^{-1/2})^{-1}\widehat L^{-1/2} U\widehat D^{-1}\\
 &= \widehat D^{-1} + \widehat D^{-1} U^\top (\widehat L- U\widehat D^{-1} U^\top)^{-1}U\widehat D^{-1}\\
&= \widehat D^{-1} \big( \widehat D + U^\top L^{-1} U \big) \widehat D^{-1}.
\end{align*}
%%%%%%%%%%%%%%%%% OLD PART %%%%%%%%%%%%%%%%%%%%%
%Therefore~\eqref{eq:zumAnderen} together with the Sherman-Morrison-Woodbury formula implies
%\begin{align}\label{SMW}
% \widehat L^{-1} = L^{-1} - L^{-1} U \big( \widehat D + U^\top L^{-1} U 
%\big)^{-1} U^\top L^{-1};
%\end{align}
%note that the matrix $U^\top L^{-1} U$ is nonnegative and hence $\widehat 
%D + U^\top L^{-1} U$ is invertible. Multiplying \eqref{SMW} from left and 
%right with $U^\top$ and $U$, respectively, we find
%\begin{align*}
% U^\top \widehat L^{-1} U & = U^\top L^{-1} U - U^\top L^{-1} U \big( \widehat D + U^\top L^{-1} U \big)^{-1} U^\top L^{-1} U \\
% & = \big( I + \widehat D \big( \widehat D + U^\top L^{-1} U \big)^{-1} 
%- I \big) U^\top L^{-1} U \\
% & = \widehat D - \widehat D \big( \widehat D + U^\top L^{-1} U \big)^{-1} \widehat D,
%\end{align*}
%which yields
%\begin{align*}
% \big( \widehat D - U^\top \widehat L^{-1} U \big)^{-1} = \widehat D^{-1} \big( \widehat D + U^\top L^{-1} U \big) \widehat D^{-1}.
%\end{align*}
%%%%%%%%%%%%%%%%% END OLD PART %%%%%%%%%%%%%%
Observe that the matrix $\widehat D - U^\top \widehat L^{-1} U$ can be obtained from $S$ by setting its $i_0$-th diagonal entry equal to $w_{e_{i_0}}$ and all further entries in the $i_0$-th row and $i_0$-th column to zero. Due to this particular structur of the $i_0$-th row and column of this matrix it follows that
\begin{align}\label{eq:Si0Inverse}
 S_{i_0}^{-1} = \left[ \widehat D^{-1} \big( \widehat D + U^\top L^{-1} 
U \big) \widehat D^{-1} \right]_{i_0},
\end{align}
where the index $i_0$ again indicates that the $i_0$-th diagonal entry is 
reset to one and all other entries in the $i_0$-th row and column are reset to zero.

{\em \underline{Step~2:}} Here we show that the $j$-th diagonal entry of~\eqref{eq:Si0Inverse} equals $\d_\cG (v_j, v_{i_0})$, the distance of the 
$j$-th boundary vertex of $\cG$ to $v_{i_0}$. Note that for $j \neq i_0$
\begin{align}\label{eq:nice}
\begin{split}
 \big\langle S_{i_0}^{-1} \e_j, \e_j \big\rangle & = \big\langle \widehat D^{-1} \big( \widehat D + U^\top L^{-1} U \big) \widehat D^{-1} \e_j, \e_j \big\rangle \\
 & = \big\langle \widehat D^{-1} \e_j, \e_j \big\rangle + \big\langle L^{-1} U \widehat D^{-1} \e_j, U \widehat D^{-1} \e_j \big\rangle,
\end{split}
\end{align}
that $\widehat D^{-1} \e_j = \frac{1}{w_{e_j}} \e_j$, where $e_j$ is the edge incident to $v_j$, and that, hence, 
\begin{align*}
 U \widehat D^{-1} \e_j = \frac{1}{w_{e_j}} U \e_j = \frac{1}{w_{e_j}} w_{e_j} \e_l
\end{align*}
if the $l$-th interior vertex $v_{k + l}$ is the one that is adjacent to $v_j$. From this,~\eqref{eq:nice} and Lemma~\ref{lem:Kirkland} we get
\begin{align*}
 \langle S_{i_0}^{-1} \e_j, \e_j \rangle & = \frac{1}{w_{e_j}} + \big\langle L^{-1} \e_l, \e_l \big\rangle = \frac{1}{w_{e_j}} +\d_\cG (v_{k + 
l}, v_{i_0}) = \d_\cG (v_j, v_{i_0}),
\end{align*}
where we have used that $L$ is the result of reducing the Laplacian on the subtree spanned by the vertex set $(\cV \setminus \partial \cG) \cup \{v_{i_0}\}$ with respect to $v_{i_0}$. This proves \eqref{eq:Hauptgewinn}. 
The latter together with Proposition \ref{prop:lengths} implies that $M_\cG$ determines the tree $\cG$ and its edge weights uniquely.
\end{proof}

\begin{remark}
Expressed in different terms, Theorem~\ref{thm:main} states that the Schur complement of the component $\widetilde L$ in the discrete Laplacian $L 
(\cG)$ determines $\cG$ and, thus, $L (\cG)$ itself uniquely if $\cG$ is a tree. It is easy to see that in general the Schur complement of a block 
does not determine the original matrix uniquely. For instance the Schur complements of the right lower corners in the block matrices  
\begin{align*}
 \begin{pmatrix}A&B\\B^\top &C\end{pmatrix} \qquad \text{and}\qquad   \begin{pmatrix} A & U B \\ (U B)^\top & U C U^\top \end{pmatrix}
\end{align*}
equal $A-B^\top C^{-1}B$ and $A-B^\top U^\top U C^{-1}U^\top U B$, respectively, and therefore they coincide for any orthogonal matrix $U$.
\end{remark}

\section{Examples}\label{sec:examples}

In this section we provide examples which show that Theorem~\ref{thm:main} does not extend, e.g., to graphs with cycles, to the Dirichlet-to-Neumann matrix on only a part of the boundary or to the so-called Weyl vector.

We start with an example showing that one cannot recover graphs with cycles uniquely from the Dirichlet-to-Neumann matrix. We point out that this example is a manifestation of the more general principle that the Dirichlet-to-Neumann matrix is invariant under so-called \emph{$\Delta$-Y-transformations}, see \cite[Section 5]{CurtInge98}.

\begin{example}\label{ex3}
Let $\cG_1$ be a graph with all edge weights equal to one that consists of a cycle with three pending edges attached to different points on the cycle, see the left-hand side of Figure~\ref{fig:ex3}. 
\begin{figure}[h]
    \centering
\begin{tikzpicture}[line cap=round,line join=round,>=triangle 45,x=1cm,y=1cm]
\clip(0.11129811797650492,1.0954025299534) rectangle (6.320509329621419,4.213798613354924);
\draw [line width=1.2pt] (0.3576694619692163,1.2426265359645343)-- (1.0005407107193824,2.008600789794519);
\draw (1.2350409989866937,2.4948929099087666) node[anchor=north west] {$\cG_1$};
\draw [line width=1.2pt] (2.7551277815476,1.2864824808348705)-- (2.0598419193444846,2.0052157900814023);
\draw [line width=1.2pt] (4.757882597292953,2.309787861142716)-- (4.770577259679596,3.64272741174015);
\draw [line width=1.2pt] (4.757882597292953,2.309787861142716)-- (5.812593541010223,1.4941447313346938);
\draw [line width=1.2pt] (4.757882597292953,2.309787861142716)-- (3.7281887736033053,1.462781651333489);
\draw (4.596786585284411,2.052063095393797) node[anchor=north west] {$\cG_2$};
\draw [line width=1.2pt] (1.545511627905529,3.9214827216061283)-- (1.5259075529710833,2.9216748999494038);
\draw [line width=1.2pt] (1.0005407107193824,2.008600789794519)-- (1.5259075529710833,2.9216748999494038);
\draw [line width=1.2pt] (1.5259075529710833,2.9216748999494038)-- (2.0598419193444846,2.0052157900814023);
\draw [line width=1.2pt] (2.0598419193444846,2.0052157900814023)-- (1.0005407107193824,2.008600789794519);
\begin{scriptsize}
\draw [fill=black] (1,2) circle (2.5pt);
\draw [fill=black] (2.0525426768880717,2) circle (2.5pt);
\draw [fill=black] (0.3576694619692163,1.2426265359645343) circle (2.5pt);
\draw [fill=black] (1.0005407107193824,2.008600789794519) circle (2.5pt);
\draw [fill=black] (2.7551277815476,1.2864824808348705) circle (2.5pt);
\draw [fill=black] (2.0598419193444846,2.0052157900814023) circle (2.5pt);
\draw [fill=black] (4.757882597292953,2.309787861142716) circle (2.5pt);
\draw [fill=black] (4.770577259679596,3.64272741174015) circle (2.5pt);
\draw [fill=black] (5.812593541010223,1.4941447313346938) circle (2.5pt);
\draw [fill=black] (3.7281887736033053,1.462781651333489) circle (2.5pt);
\draw [fill=black] (1.545511627905529,3.9214827216061283) circle (2.5pt);
\draw [fill=black] (1.5259075529710833,2.9216748999494038) circle (2.5pt);
\end{scriptsize}
\end{tikzpicture}
    \caption{The equilateral graphs from Example~\ref{ex3}. If $\cG_1$ has edge weights $1$ and $\cG_2$ has edge weights $3/4$ then the two graphs 
have the same Dirichlet-to-Neumann-matrix.}
    \label{fig:ex3}
\end{figure}
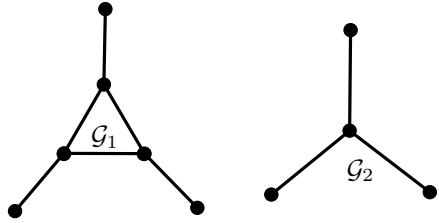
The discrete Laplacian of $\cG_1$ is then given by
\[
 L(\cG_1)=\begin{pmatrix}1&0&0&-1&0&0\\ 0&1&0&0&-1&0\\0&0&1&0&0&-1\\-1&0&0&3&-1&-1\\ 0&-1&0&-1&3&-1\\0&0&-1&-1&-1&3 \end{pmatrix}.
\]
We may use the identity \eqref{DtoN_schur} to calculate the Dirichlet-to-Neumann matrix for $\cG_1$ and obtain
\[
 M_{\cG_1} = I_3 - I_3 \begin{pmatrix}3&-1&-1\\ -1&3&-1\\-1&-1&3\end{pmatrix}^{-1} I_3=\frac{1}{4}\begin{pmatrix}2&-1&-1\\-1&2&-1\\-1&-1&2\end{pmatrix}.
\]
Let now $\cG_2$ be a 3-star with all edge weights equal to $3/4$, see the 
right-hand side of Figure~\ref{fig:ex3}. Then the discrete Laplacian of $\cG_2$ equals 
\[
 L(\cG_2)=\frac{3}{4}\begin{pmatrix}1&0&0&-1\\0&1&0&-1\\0&0&1&-1\\-1&-1&-1&3\end{pmatrix}
\]
and, hence, the Dirichlet-to-Neumann matrix on $\cG_2$ is given by
\[
 M_{\cG_2} = \frac{3}{4} I_3 - \frac{1}{4} \begin{pmatrix} 1 \\ 1 \\ 1 \end{pmatrix}\begin{pmatrix} 1 & 1 & 1 \end{pmatrix} = M_{\cG_1}.
\]
Therefore the Dirichlet-to-Neumann matrix alone cannot determine a weighted graph uniquely if cycles are allowed.
\end{example}

Another consequence of the previous example is that it is not possible either to recover the Betti number (i.e.\ the number of independent cycles) 
from the knowledge of $M_\cG$ only.

The next example shows that the Dirichlet-to-Neumann matrix for a proper subset of the leaves does not determine a weighted tree uniquely.

\begin{example}\label{ex:Soundso}
Consider a 3-star $\cG_1$ with edge weights one and a path graph $\cG_2$ consisting of a single edge of weight $1/2$; cf.\ Figure~\ref{fig:exSoundso}. 
\begin{figure}[h]
\centering
\begin{tikzpicture}[line cap=round,line join=round,>=triangle 45,x=1cm,y=1cm]
% star graph
\draw [line width=1.2pt] (4.757882597292953,2.309787861142716)-- (4.770577259679596,3.64272741174015);
\draw [line width=1.2pt] (4.757882597292953,2.309787861142716)-- (5.812593541010223,1.4941447313346938);
\draw [line width=1.2pt] (4.757882597292953,2.309787861142716)-- (3.7281887736033053,1.462781651333489);
\draw (4.583333333333333,1.9044943820224718) node[anchor=north west] {$\cG_1$};
% path graph
\draw [line width=1.2pt] (8,2.5)-- (12,2.5);
\draw (9.783333333333333,2.3044943820224718) node[anchor=north west] {$\cG_2$};
\begin{scriptsize}
\draw [fill=black] (4.770577259679596,3.64272741174015) circle (2.5pt);
\draw [fill=black] (5.812593541010223,1.4941447313346938) circle (2.5pt);
\draw [fill=black] (3.7281887736033053,1.462781651333489) circle (2.5pt);
\draw [fill=black] (8,2.5) circle (2.5pt);
\draw [fill=black] (12,2.5) circle (2.5pt);
\end{scriptsize}
\end{tikzpicture}
    \caption{The graphs from Example~\ref{ex:Soundso}. If $\cG_1$ has edge weights one and $\cG_2$ has weight $1/2$ then the Dirichlet-to-Neumann matrix of $\cG_1$ with respect to only two boundary vertices coincides with $M_{\cG_2}$.}
    \label{fig:exSoundso}
\end{figure}
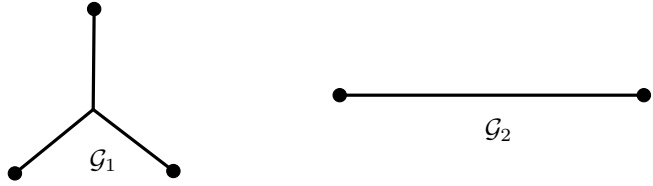
The discrete Laplacians of the corresponding weighted trees $\cG_1$ and $\cG_2$ are
\[
 L (\cG_1) = \begin{pmatrix}1&0&0&-1\\0&1&0&-1\\0&0&1&-1\\-1&-1&-1&3 \end{pmatrix} \quad \text{and} \quad L(\cG_2)=\begin{pmatrix}1/2&-1/2\\-1/2&1/2\end{pmatrix} = M_{\cG_2}.
\]
The Dirichlet-to-Neumann matrix of $\cG_1$ with respect to only two boundary vertices (i.e.\ the $2 \times 2$-matrix defined as Schur complement of the block corresponding to all vertices except the first two leaves in $L (\cG_1)$) is given by
\[
 \begin{pmatrix}1&0\\0&1 \end{pmatrix}-\begin{pmatrix} 0 & 1 \\ 0 & 1 \end{pmatrix} \begin{pmatrix} 1 & -1 \\ -1 & 3 \end{pmatrix}^{-1} \begin{pmatrix} 0 & 0 \\ 1 & 1 \end{pmatrix} = \begin{pmatrix}1/2&-1/2\\-1/2&1/2\end{pmatrix},
\]
which coincides with $M_{\cG_2}$.
\end{example}

In some places in the literature inverse problems for quantum trees (see Section \ref{sec:QG} below) were solved under the assumption that not the 
whole $\lambda$-dependent Dirichlet-to-Neumann matrix but only its diagonal, the so-called {\em Weyl vector} is available, see, e.g.~\cite{AK08,FY07,Y05}. The following example shows that the knowledge of only the diagonal of $M_{\cG}$ does not determine the weighted tree $\cG$ uniquely.

\begin{example}
\label{ex4}
Consider, on the one hand, the star $\cG_1$ consisting of four edges of weight $1$ each, see the left-hand side of Figure~\ref{fig:ex4}. 
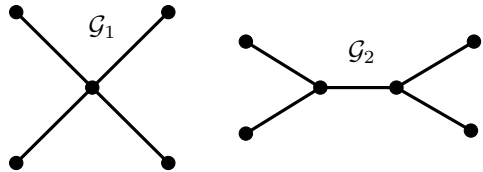
\begin{figure}[h]
    \centering
 \begin{tikzpicture}[line cap=round,line join=round,>=triangle 45,x=1cm,y=1cm]
\clip(-6,-1.2) rectangle (2,2);
\draw [line width=1.2pt] (-4,0)-- (-5,1);
\draw [line width=1.2pt] (-4,0)-- (-3,1);
\draw [line width=1.2pt] (-4,0)-- (-3,-1);
\draw [line width=1.2pt] (-4,0)-- (-5,-1);
\draw [line width=1.2pt] (-1,0)-- (-1.98,0.61);
\draw [line width=1.2pt] (-1,0)-- (0,0);
\draw [line width=1.2pt] (-1,0)-- (-1.98,-0.61);
\draw [line width=1.2pt] (0,0)-- (1.02,0.61);
\draw [line width=1.2pt] (0,0)-- (0.98,-0.57);
\draw (-4.171779141104294,1.07592867756315) node[anchor=north west] {$\cG_1$};
\draw (-0.7607361963190185,0.7451708766716196) node[anchor=north west] {$\cG_2$};
\draw (-4.441717791411043,-0.6567607726597325) node[anchor=north west] {};
\draw (-4.7975460122699385,0.3714710252600297) node[anchor=north west] {};
\draw (-3.337423312883436,0.3655274888558692) node[anchor=north west] {};
\draw (-1.8159509202453987,0.29420505200594355) node[anchor=north west] 
{};
\draw (0.7484662576687117,0.30609212481426445) node[anchor=north west] {};
\draw (0.7852760736196319,-0.1337295690936107) node[anchor=north west] {};
\draw (-1.852760736196319,-0.11589895988112926) node[anchor=north west] 
{};
\begin{scriptsize}
\draw [fill=black] (-4,0) circle (2.5pt);
\draw [fill=black] (-5,1) circle (2.5pt);
\draw [fill=black] (-3,1) circle (2.5pt);
\draw [fill=black] (-5,-1) circle (2.5pt);
\draw [fill=black] (-3,-1) circle (2.5pt);
\draw [fill=black] (-1.98,0.61) circle (2.5pt);
\draw [fill=black] (-1,0) circle (2.5pt);
\draw [fill=black] (-1.98,-0.61) circle (2.5pt);
\draw [fill=black] (0,0) circle (2.5pt);
\draw [fill=black] (1.02,0.61) circle (2.5pt);
\draw [fill=black] (0.98,-0.57) circle (2.5pt);
\end{scriptsize}
\end{tikzpicture}
    \caption{The two graphs from Example \ref{ex4}. If $\cG_1$ has edge weights one and $\cG_2$ has edge weights $6/5$ then $M_{\cG_1}$ and $M_{\cG_2}$ have the same diagonal.}
    \label{fig:ex4}
\end{figure}
Then the corresponding weighted star graph $\cG_1$ has edge weights $w_e = 1$ for each edge $e$ and its discrete Laplacian is given by
\begin{align*}
 L (\cG_1) = \begin{pmatrix} 1 & 0 & 0 & 0 & - 1 \\ 0 & 1 & 0 & 0 & - 1 
\\ 0 & 0 & 1 & 0 & - 1 \\ 0 & 0 & 0 & 1 & - 1 \\ - 1 & - 1 & - 1 & - 1 & 4 \end{pmatrix}.
\end{align*}
Using \eqref{DtoN_schur} we get the corresponding Dirichlet-to-Neumann matrix
\begin{align*}
 M_{\cG_1} = I_4 - \frac{1}{4} \begin{pmatrix} 1 & 1 & 1 & 1 \\ 1 & 1 & 
1 & 1 \\ 1 & 1 & 1 & 1 \\ 1 & 1 & 1 & 1 \end{pmatrix} = \frac{1}{4} \begin{pmatrix} 3 & - 1 & - 1 & - 1 \\ - 1 & 3 & - 1 & - 1 \\ - 1 & - 1 & 3 & - 1 \\ - 1 & - 1 & - 1 & 3 \end{pmatrix}.
\end{align*}
In particular, each diagonal entry of $M_{\cG_1}$ equals $3/4$.

On the other hand, for the symmetric ``double star'' graph $\cG_2$ on the 
right-hand side of Figure~\ref{fig:ex4} with four boundary edges of weight $a$ and one interior edge of weight $b$ the discrete Laplacian equals
\begin{align*}
 L (\cG_2) = \begin{pmatrix} a & 0 & 0 & 0 & - a & 0 \\ 0 & a & 0 & 0 & 
- a & 0 \\ 0 & 0 & a & 0 & 0 & - a \\ 0 & 0 & 0 & a & 0 & - a \\ - a & - a & 0 & 0 & 2 a + b & - b \\ 0 & 0 & - a & - a & - b & 2 a + b \end{pmatrix}
\end{align*}
and, thus, by \eqref{DtoN_schur},
\begin{align*}
 M_{\cG_2} = a I_4 - \frac{1}{4 a (a + b)} \begin{pmatrix} a^2 (2 a + b) & a^2 (2 a + b) & a^2 b & a^2 b \\ a^2 (2 a + b) & a^2 (2 a + b) & a^2 b & a^2 b \\ a^2 b & a^2 b & a^2 (2 a + b) & a^2 (2 a + b) \\ a^2 b & a^2 
b & a^2 (2 a + b) & a^2 (2 a + b) \end{pmatrix}.
\end{align*}
In particular, each diagonal entry equals $a - \frac{a (2 a + b)}{4 (a + b)}$. Setting, e.g., $a = b = 6/5$ we get all diagonal entries equal to $3/4$ and thus the diagonals of $M_{\cG_1}$ and $M_{\cG_2}$ coincide in this case while $\cG_1$ and $\cG_2$ differ from each other.
\end{example}

\section{A remark on the Dirichlet-to-Neumann matrix for quantum graphs}
\label{sec:QG}

We have mentioned earlier that the matrix $M_\cG$ also has an interpretation as Dirichlet-to-Neumann matrix of a quantum graph. In this short section we explain this in more detail. Related to any weighted graph $\cG$ with edge weights $w_e$, $e \in \cE$, we denote by $\Gamma$ the corresponding metric graph with edge lengths
\begin{align*}
 L (e) := \frac{1}{w_e}, \quad e \in \cE;
\end{align*}
we identify each edge $e$ of $\Gamma$ with the interval $[0, L (e)]$ and obtain from this parametrization a natural metric on $\Gamma$. For a function $f : \Gamma \to \C$ we denote by $f_e := f |_e$ its restriction to 
the edge $e$. For any function $f$ on $\Gamma$ such that $f_e \in H^2 (0, 
L (e))$ for each $e$ we define the normal derivative $\partial f (v)$ at any vertex $v$ as
\[
 \partial f (v) := \sum_{e \in \cE (v)} \frac{\d}{\d x} f_e (v),
\]
where the derivatives are taken in the direction towards the vertex.  For a chosen boundary $\partial \cG = \{v_1, \dots, v_k\}$ we may then define the Dirichlet-to-Neumann matrix for the Laplacian with standard vertex conditions to be the matrix $M_\Gamma \in \R^{k \times k}$ such that
\begin{align*}
 M_\Gamma \begin{pmatrix} f (v_1) \\ \vdots \\ f (v_k) \end{pmatrix} = \begin{pmatrix} \partial f (v_1) \\ \vdots \\ \partial f (v_k) \end{pmatrix}
\end{align*}
holds for any function $f$ on $\Gamma$ such that $f_e \in H^2 (0, L (e))$ 
holds for each $e \in \cE$, $f$ is continuous and fulfills the Kirchhoff condition $\partial f (v) = 0$ at each vertex $v \in \cV \setminus \partial \cG$ and satisfies $f_e'' = 0$ constantly inside every edge $e$. It can be shown that $M_\Gamma$ is well-defined, see, e.g., \cite[Section 3.5]{BK13}.

However, the matrix $M_\Gamma$ coincides with $M_\cG$ as defined in Definition \ref{def:DN}. Indeed, let $f$ be a function with the mentioned properties, e.g.\ it is harmonic on every edge and satisfies continuity--Kirchhoff (also called standard) vertex conditions at the interior vertices. It is clear that such $f$ is linear on every edge; if we take an arbitrary $v_i \in \cV$ and assume for simplity that all edges of $\Gamma$ to which $v_i$ is incident are parametrized such that they terminate at $v_i$, then for any such edge $e$ and the corresponding originating vertex $v_j$,
\begin{align*}
 f_e (x) = \frac{x}{L (e)} f (v_i) + \frac{L (e) - x}{L (e)} f (v_j),
\end{align*}
and thus
\begin{align*}
 f_e' (v_i) = \frac{1}{L (e)} \big( f_e (v_i) - f_e (w) \big) = w_e \big( f (v) - f (w) \big)
\end{align*}
by continuity. Let now $u := f |_V$ denote the restriction of $f$ to the vertices. Then for each $i$,
\begin{align*}
 \big(L (\cG) u \big) (v_i) = \sum_{e \in \cE (v_i) \cap \cE (v_j)} w_e 
\big( f (v_i) - f (v_j) \big) = \sum_{e \in \cE (v_i)} f_e' (v_i).
\end{align*}
This is zero if $v_i \in \cV \setminus \partial \cG$. Hence, for $v_i \in 
\partial \cG$,
\begin{align*}
 (M_\cG f |_{\partial \cG}) (v_i) = \frac{\partial u}{\partial \rm n} (v_i) = \sum_{e \in \cE (v_i)} f_e' (v_i) = \partial f (v_i).
\end{align*}
Thus $M_\cG = M_\Gamma$. Consequently, Theorem \ref{thm:main} may be rephrased as follows; here we again choose $\partial \cG$ to be the set of all vertices of degree one.

\begin{theorem}
The Dirichlet-to-Neumann matrix $M_\Gamma$ for the Laplacian with continuity-Kirchhoff vertex conditions determines the metric tree $\Gamma$ uniquely up to vertices of degree two.
\end{theorem}

%\begin{acknowledgements}\label{ackref}
%The \verb"acknowledgements" environment may be used to acknowledge
%indebtedness to colleagues, host institutions and referees. Accounts
%of grants and financial support should be made as a footnote on the
%title page using the \verb"\extraline{}" command in the preamble.
%\end{acknowledgements}


\begin{thebibliography}{9}% Replace 9 by 99 if 10 or more references
%
% Please note the use of "\and" between author names below
%

\bibitem{AIM18} {\sc K.\ Ando, H.\ Isozaki, \text{and} H.\ Morioka}, `Inverse scattering for Schr\"odinger operators on perturbed lattices', \emph{Ann.\ Henri Poincar\'e} 19 (2018), no. 11, 3397--3455.

\bibitem{AK08} {\sc S.~Avdonin and P.~Kurasov}, `Inverse problems for quantum trees', \emph{Inverse Probl.\ Imaging} 2 (2008)  1--21.

\bibitem{AP06} {\sc K.~Astala and L.~P\"{a}iv\"{a}rinta}, `Calder\'on's inverse conductivity problem in the plane', \emph{Ann. of Math.} (2) 
163 (2006) 265--299.


\bibitem{BK13} {\sc G.~Berkolaiko and P.~Kuchment},  \emph{Introduction to quantum graphs}  (AMS, Providence, RI, 2013). %\emph{Math.\ Surveys and Monographs vol.~186},

\bibitem{BGG16} {\sc J.\ Boyer, J.\,J.\ Garzella, and F.\ Guevara Vasquez}, `On the solvability of the discrete conductivity and Schrödinger inverse problems', \emph{SIAM J.\ Appl.\ Math.} 76 (2016), no. 3, 1053--1075. 

\bibitem{BW05} {\sc B.\,M.~Brown and R.~Weikard}, `A Borg--Levinson 
theorem for trees', \emph{Proc.\ R.\ Soc.\ Lond.\ Ser.~A Math.\ Phys.\ Eng.\ Sci.} 461 (2005) 3231--3243.

\bibitem{C80} {\sc A.\,P.~Calder\'on}, `On an inverse boundary value 
problem', \emph{Seminar on Numerical Analysis and its Applications to Continuum Physics (Rio de Janeiro, 1980)} 65--73. %, Soc. Brasil. Mat., Rio de Janeiro, 1980.

%%%%fand ich zu lang %%%%%%%%%%%
\bibitem{C97} {\sc F.\,R.\,K.\ Chung}, \emph{Spectral graph theory} (AMS, Providence, RI, 1997).

%\bibitem{C97} {\bibname F.\,R.\,K.\ Chung}, \emph{Spectral graph theory} 
(CBMS Regional Conference Series in Mathematics, 92. Published for the Conference Board of the Mathematical Sciences, Washington, DC; by the American Mathematical Society, Providence, RI, 1997).


\bibitem{BereChun05} {\sc S.-Y.~Chung and C.\,A.~Berenstein}, `$\omega$-harmonic functions and inverse conductivity problems on networks', \emph{SIAM J. Appl.\  Math.} 65 (2005) 1200--1226. 

\bibitem{CGHS17} {\sc F.\,J.\ Chung, A.\,C.\ Gilbert, J.\,G.\ Hoskins, and J.\,C.\ Schotland}, `Optical tomography on graphs', \emph{Inverse 
Problems} 33 (2017), no. 5, 055016, 21 pp.

\bibitem{CW09} {\sc S.~Currie and B.\,A.~Watson}, `The M-matrix inverse problem for the Sturm--Liouville equation on graphs', \emph{Proc.\ Roy.\ Soc.\ Edinburgh Sect.\ A} 139 (2009) 775--796. 

\bibitem{CurtMooe94} {\sc E.\,B.~Curtis, E.~Mooers, and J.\,A.~Morrow}, `Finding the conductors in circular networks from boundary measurements', \emph{ESAIM: Mathematical Modelling and Numerical Analysis} 28 (1994) 781--814. 

\bibitem{CurtMorr90} {\sc E.\,B.~Curtis and J.\,A.~Morrow}, `Determining the resistors in a network', \emph{SIAM J.\ Appl.\ Math.} 50 (1990) 
918--930. 

\bibitem{CurtMorr91} {\sc E.\,B.~Curtis and J.\,A.~Morrow}, `The Dirichlet to Neumann map for a resistor network', \emph{SIAM J.\ Appl.\ Math.} 51 (1991) 1011--1029. 

\bibitem{CurtInge98} {\sc E.\,B.~Curtis, D.~Ingerman, and J.\,A.~Morrow}, `Circular planar graphs and resistor networks', \emph{Linear Algebra Appl.} 283 (1998) 115--150. 


\bibitem{EK11} {\sc M.~Enerb\"ack and P.~Kurasov}, `Aharonov-Bohm ring touching a quantum wire: how to model it and to solve the inverse problem', \emph{Rep.\ Math.\ Phys.} 68 (2011)  271--287.


\bibitem{FY07} {\sc G.~Freiling and V.~Yurko}, `Inverse problems for Sturm--Liouville operators
on noncompact trees', \emph{Results Math.} 50 (2007) 195--212.

\bibitem{HakiYau65} {\sc S.\,L.~Hakimi and S.\,S.~Yau}, `Distance matrix of a graph and its realizability', \emph{Q.\  Appl.\ Math.} 22 (1965) 305--317.

\bibitem{Kirkland1996} {\sc S.\ Kirkland and B.\ Shader}, `Characteristic vertices of weighted trees via {P}erron values',
\emph{Linear Multilinear A.} 40 (1996) 311--325.


\bibitem{KleiRand93} {\sc D.\,J.\ Klein and M.\ Randi\'{c}}, `Resistance distance',
\emph{J.\ Math.\ Chem.} 12 (1993) 81--95.

\bibitem{K13} {\sc P.~Kurasov}, `Inverse scattering for lasso graph', \emph{J.\ Math.\ Phys.} 54 (2013) 042103. 

\bibitem{N88} {\sc A.~Nachman}, `Reconstructions from boundary measurements', \emph{Ann. of Math.} (2) 128 (1988) 531--576.

\bibitem{N96} {\sc A.~Nachman}, `Global uniqueness for a two-dimensional inverse boundary value problem', \emph{Ann. of Math.} (2) 143 (1996) 
71--96.

\bibitem{R15} {\sc J.~Rohleder}, `Recovering a quantum graph spectrum from vertex 
data', \emph{J.\ Phys.~A} 48 (2015) 165202.

\bibitem{Ruec11} {\sc R.~Rueckriemen}, `Quantum graphs and their spectra', PhD Thesis, Dartmouth College, Hanover, New Hampshire, 2011.

\bibitem{SU87} {\sc J.~Sylvester and G.~Uhlmann}, `A global uniqueness theorem for an inverse boundary value problem', \emph{Ann. of Math.} (2) 125 (1987) 153--169.

\bibitem{Y05} {\sc V.~Yurko}, `Inverse spectral problems for Sturm-Liouville operators on graphs', \emph{Inverse Problems} 21 (2005) 1075--1086.

\bibitem{Y12} {\sc V.~Yurko}, `Inverse spectral problems for arbitrary order differential operators on noncompact trees', {\em J.~Inverse Ill-Posed Probl.} 20 (2012) 111--131.

%\bibitem{Goddard}
%{\bibname P. Goddard, A. Kent \and D. I. Olive}, `Unitary
%representations of the Virasoro and Supervirasoro algebras', {\em
%Comm. Math. Phys. }103 (1986) 105.
%
%\bibitem{Lamport}
% {\bibname L. Lamport},
% {\em\LaTeX: A document preparation system {\rm (}updated for
% \LaTeXe{\rm)}} (Addison-Wesley, New York, 1994).
%
%\bibitem{Lance}
% {\bibname E. C. Lance \and A. Paolucci},
% `Conjugation in braided C$^*$-categories and
% orthogonal quantum groups',
% {\em J. Math. Phys. }41 (2000) 2383--2394.
%
%\bibitem{Stuart}
%{\bibname J. T. Stuart}, `Mathematics applied in fluid motion',
%{\em Quart. Appl. Math. }56 (1998) 787--796.
%
%\bibitem{PRL}
% {\bibname
%T.~Prokopec, O.~T\"ornkvist \and R.~P.~Woodard}, `Photon mass
%  from inflation', \emph{Phys. Rev. Lett. }89 (2002) 101301.
\end{thebibliography}
\end{document}